\documentclass[10pt,letterpaper,conference]{IEEEtran}

\usepackage{cite}
\usepackage{amsmath,amssymb,amsfonts}
\usepackage{algorithmic}
\usepackage{graphicx}
\usepackage{textcomp}
\usepackage{xcolor}
\usepackage{comment}
\usepackage[detect-all=true,per-mode=repeated-symbol,per-symbol =/]{siunitx}
\usepackage[acronym]{glossaries}
\usepackage[hidelinks]{hyperref}
\usepackage[capitalise]{cleveref}
\usepackage{float}
\usepackage{subcaption}
\usepackage{booktabs}
\usepackage{threeparttable}
\usepackage{pifont}
\usepackage{lipsum}
\usepackage{makecell}
\usepackage{colortbl}

\usepackage[paperwidth=10in, paperheight=12.5in, top=1.5in, bottom=1.75in, left=1.0in, right=1.0in]{geometry}

\definecolor{ieee-bright-dblue-100}{rgb}{0.0, 0.3828, 0.6055}
\definecolor{ieee-bright-dblue-80}{rgb}{0.0, 0.4883, 0.6797}
\definecolor{ieee-bright-dblue-60}{rgb}{0.3633, 0.6094, 0.7617}
\definecolor{ieee-bright-dblue-40}{rgb}{0.5898, 0.7383, 0.8398}
\definecolor{ieee-bright-dblue-20}{rgb}{0.8906, 0.8984, 0.9219}
\definecolor{ieee-bright-red-100}{rgb}{0.7266, 0.0469, 0.1836}
\definecolor{ieee-bright-red-80}{rgb}{0.832, 0.3164, 0.3281}
\definecolor{ieee-bright-red-60}{rgb}{0.8906, 0.4922, 0.4805}
\definecolor{ieee-bright-red-40}{rgb}{0.9336, 0.6562, 0.6406}
\definecolor{ieee-bright-red-20}{rgb}{0.9688, 0.8203, 0.8125}
\definecolor{ieee-bright-orange-100}{rgb}{0.9961, 0.6367, 0.0}
\definecolor{ieee-bright-orange-80}{rgb}{0.9844, 0.6953, 0.3125}
\definecolor{ieee-bright-orange-60}{rgb}{0.9883, 0.7695, 0.4844}
\definecolor{ieee-bright-orange-40}{rgb}{0.9922, 0.8359, 0.6562}
\definecolor{ieee-bright-orange-20}{rgb}{0.9961, 0.9219, 0.8164}
\definecolor{ieee-bright-yellow-100}{rgb}{0.9961, 0.8164, 0.0}
\definecolor{ieee-bright-yellow-80}{rgb}{0.9961, 0.8477, 0.2148}
\definecolor{ieee-bright-yellow-60}{rgb}{0.9961, 0.875, 0.4492}
\definecolor{ieee-bright-yellow-40}{rgb}{0.9961, 0.9062, 0.6328}
\definecolor{ieee-bright-yellow-20}{rgb}{0.9961, 0.9531, 0.8125}
\definecolor{ieee-bright-lgreen-100}{rgb}{0.4688, 0.7422, 0.125}
\definecolor{ieee-bright-lgreen-80}{rgb}{0.5742, 0.7852, 0.332}
\definecolor{ieee-bright-lgreen-60}{rgb}{0.6875, 0.8398, 0.5039}
\definecolor{ieee-bright-lgreen-40}{rgb}{0.793, 0.8906, 0.6641}
\definecolor{ieee-bright-lgreen-20}{rgb}{0.8945, 0.9414, 0.8281}
\definecolor{ieee-bright-dgreen-100}{rgb}{0.0, 0.5156, 0.2383}
\definecolor{ieee-bright-dgreen-80}{rgb}{0.1641, 0.6055, 0.3867}
\definecolor{ieee-bright-dgreen-60}{rgb}{0.3906, 0.6953, 0.5234}
\definecolor{ieee-bright-dgreen-40}{rgb}{0.6094, 0.8008, 0.6719}
\definecolor{ieee-bright-dgreen-20}{rgb}{0.8047, 0.8945, 0.8359}
\definecolor{ieee-bright-purple-100}{rgb}{0.5938, 0.1133, 0.5898}
\definecolor{ieee-bright-purple-80}{rgb}{0.6992, 0.3281, 0.668}
\definecolor{ieee-bright-purple-60}{rgb}{0.7812, 0.4961, 0.7461}
\definecolor{ieee-bright-purple-40}{rgb}{0.8555, 0.6602, 0.8281}
\definecolor{ieee-bright-purple-20}{rgb}{0.9219, 0.8281, 0.9023}
\definecolor{ieee-bright-lblue-100}{rgb}{0.0, 0.6094, 0.6484}
\definecolor{ieee-bright-lblue-80}{rgb}{0.0, 0.6797, 0.7188}
\definecolor{ieee-bright-lblue-60}{rgb}{0.2109, 0.75, 0.7812}
\definecolor{ieee-bright-lblue-40}{rgb}{0.5469, 0.8242, 0.8438}
\definecolor{ieee-bright-lblue-20}{rgb}{0.7695, 0.918, 0.9219}
\definecolor{ieee-bright-cyan-100}{rgb}{0.0, 0.707, 0.8828}
\definecolor{ieee-bright-cyan-80}{rgb}{0.0, 0.7227, 0.9453}
\definecolor{ieee-bright-cyan-60}{rgb}{0.2656, 0.7812, 0.957}
\definecolor{ieee-bright-cyan-40}{rgb}{0.5547, 0.8438, 0.9688}
\definecolor{ieee-bright-cyan-20}{rgb}{0.7773, 0.9141, 0.9805}
\definecolor{ieee-bright-white-100}{rgb}{0.9961, 0.9961, 0.9961}
\definecolor{ieee-bright-white-80}{rgb}{0.9961, 0.9961, 0.9961}
\definecolor{ieee-bright-white-60}{rgb}{0.9961, 0.9961, 0.9961}
\definecolor{ieee-bright-white-40}{rgb}{0.9961, 0.9961, 0.9961}
\definecolor{ieee-bright-white-20}{rgb}{0.9961, 0.9961, 0.9961}
\definecolor{ieee-dark-red-100}{rgb}{0.5234, 0.1211, 0.2539}
\definecolor{ieee-dark-red-80}{rgb}{0.6445, 0.2812, 0.3828}
\definecolor{ieee-dark-red-60}{rgb}{0.7422, 0.4727, 0.5234}
\definecolor{ieee-dark-red-40}{rgb}{0.832, 0.6445, 0.6758}
\definecolor{ieee-dark-red-20}{rgb}{0.918, 0.8203, 0.832}
\definecolor{ieee-dark-orange-100}{rgb}{0.9062, 0.4648, 0.1328}
\definecolor{ieee-dark-orange-80}{rgb}{0.9648, 0.5664, 0.3164}
\definecolor{ieee-dark-orange-60}{rgb}{0.9766, 0.6758, 0.4805}
\definecolor{ieee-dark-orange-40}{rgb}{0.9844, 0.7773, 0.6523}
\definecolor{ieee-dark-orange-20}{rgb}{0.9922, 0.8789, 0.8125}
\definecolor{ieee-dark-yellow-100}{rgb}{0.9961, 0.7773, 0.1719}
\definecolor{ieee-dark-yellow-80}{rgb}{0.9961, 0.8086, 0.375}
\definecolor{ieee-dark-yellow-60}{rgb}{0.9961, 0.875, 0.4492}
\definecolor{ieee-dark-yellow-40}{rgb}{0.9961, 0.8984, 0.6875}
\definecolor{ieee-dark-yellow-20}{rgb}{0.9961, 0.9453, 0.8438}
\definecolor{ieee-dark-lgreen-100}{rgb}{0.3945, 0.5508, 0.0938}
\definecolor{ieee-dark-lgreen-80}{rgb}{0.5078, 0.6289, 0.293}
\definecolor{ieee-dark-lgreen-60}{rgb}{0.6367, 0.7188, 0.4688}
\definecolor{ieee-dark-lgreen-40}{rgb}{0.7539, 0.8047, 0.6367}
\definecolor{ieee-dark-lgreen-20}{rgb}{0.875, 0.9023, 0.8125}
\definecolor{ieee-dark-dgreen-100}{rgb}{0.0, 0.3867, 0.2539}
\definecolor{ieee-dark-dgreen-80}{rgb}{0.1836, 0.5, 0.3906}
\definecolor{ieee-dark-dgreen-60}{rgb}{0.3984, 0.6172, 0.5273}
\definecolor{ieee-dark-dgreen-40}{rgb}{0.5938, 0.7422, 0.6758}
\definecolor{ieee-dark-dgreen-20}{rgb}{0.793, 0.8711, 0.8359}
\definecolor{ieee-dark-purple-100}{rgb}{0.4648, 0.1445, 0.5117}
\definecolor{ieee-dark-purple-80}{rgb}{0.5898, 0.3242, 0.6016}
\definecolor{ieee-dark-purple-60}{rgb}{0.6914, 0.4883, 0.6953}
\definecolor{ieee-dark-purple-40}{rgb}{0.7969, 0.6523, 0.793}
\definecolor{ieee-dark-purple-20}{rgb}{0.8945, 0.8203, 0.8945}
\definecolor{ieee-dark-cyan-100}{rgb}{0.0, 0.4492, 0.4648}
\definecolor{ieee-dark-cyan-80}{rgb}{0.0, 0.5469, 0.5664}
\definecolor{ieee-dark-cyan-60}{rgb}{0.3047, 0.6602, 0.668}
\definecolor{ieee-dark-cyan-40}{rgb}{0.5586, 0.7695, 0.7734}
\definecolor{ieee-dark-cyan-20}{rgb}{0.7734, 0.8789, 0.8789}
\definecolor{ieee-dark-dblue-100}{rgb}{0.0, 0.1562, 0.332}
\definecolor{ieee-dark-dblue-80}{rgb}{0.1797, 0.3008, 0.4609}
\definecolor{ieee-dark-dblue-60}{rgb}{0.3828, 0.4609, 0.5859}
\definecolor{ieee-dark-dblue-40}{rgb}{0.5781, 0.6289, 0.7188}
\definecolor{ieee-dark-dblue-20}{rgb}{0.7852, 0.8047, 0.8555}
\definecolor{ieee-dark-grey-100}{rgb}{0.457, 0.4688, 0.4805}
\definecolor{ieee-dark-grey-80}{rgb}{0.5625, 0.5625, 0.5742}
\definecolor{ieee-dark-grey-60}{rgb}{0.6641, 0.6641, 0.6758}
\definecolor{ieee-dark-grey-40}{rgb}{0.7734, 0.7695, 0.7773}
\definecolor{ieee-dark-grey-20}{rgb}{0.8789, 0.8828, 0.8828}
\definecolor{ieee-dark-black-100}{rgb}{0.0, 0.0, 0.0}
\definecolor{ieee-dark-black-80}{rgb}{0.3438, 0.3477, 0.3555}
\definecolor{ieee-dark-black-60}{rgb}{0.5, 0.5078, 0.5195}
\definecolor{ieee-dark-black-40}{rgb}{0.6523, 0.6602, 0.6719}
\definecolor{ieee-dark-black-20}{rgb}{0.8164, 0.8242, 0.8281}

\newcommand{\gf}{GlobalFoundries}
\newcommand{\socc}{Occamy}
\newcommand{\sflo}{Ramora}
\newcommand{\sogo}{Ogopogo}

\DeclareSIUnit\GE{GE}
\DeclareSIUnit\kGE{\kilo\GE}
\DeclareSIUnit\MGE{\mega\GE}
\DeclareSIUnit{\x}{\!\ensuremath{\times}}
\DeclareSIUnit{\percent}{\!\%}
\DeclareSIUnit\bit{b}
\DeclareSIUnit\flop{FLOP}
\DeclareSIUnit\pin{pin}
\DeclareSIUnit\hop{hop}
\DeclareSIUnit\tok{tok}
\DeclareSIUnit\dash{\text{-}}
\ifx\showtodo\undefined
    \newcommand{\todo}[1]{{TODO: #1}}
    
\else
    \newcommand{\todo}[1]{{\textcolor{red}{TODO: #1}}}
    
\fi

\ifx\showfeedback\undefined
    \newcommand{\lb}[1]{#1}
\else
    \newcommand{\lb}[1]{{\textcolor{ieee-bright-lblue-100}{#1}}}
\fi

\ifx\showcuts\undefined
    \newcommand{\cut}[1]{}
\else
    \newcommand{\cut}[1]{{\textcolor{ieee-dark-red-100}{#1}}}
\fi

\def\refbls{0.9757}
\renewcommand{\baselinestretch}{\contentbls}

\newacronym{ai}{AI}{artificial intelligence}
\newacronym{d2d}{D2D}{die-to-die}
\newacronym{eda}{EDA}{electronic design automation}
\newacronym{fpu}{FPU}{floating-point unit}
\newacronym{gpu}{GPU}{graphics processing unit}
\newacronym{cpu}{CPU}{central processing unit}
\newacronym{hpc}{HPC}{high-performance computing}
\newacronym{ml}{ML}{machine learning}
\newacronym[longplural={networks-on-chip}]{noc}{NoC}{network-on-chip}
\newacronym{pdk}{PDK}{process design kit}
\newacronym{phy}{PHY}{physical interface}
\newacronym{rtl}{RTL}{register transfer level}
\newacronym{soa}{SoA}{state-of-the-art}
\newacronym{spm}{SPM}{scratchpad memory}
\newacronym{su}{SU}{streaming unit}
\newacronym{dma}{DMA}{direct memory access}
\newacronym{isa}{ISA}{instruction set architecture}
\newacronym{simd}{SIMD}{single-instruction, multiple-data}
\newacronym{fma}{FMA}{fused multiply-accumulate}
\newacronym{axi4}{AXI4}{advanced eXtensible interface 4}
\newacronym{iotlb}{IOTLB}{IO translation lookaside buffer}
\newacronym{ddr}{DDR}{double-data-rate}
\newacronym{ip}{IP}{intellectual property}
\newacronym{gemm}{GEMM}{general matrix multiply}
\newacronym{ate}{ATE}{automatic test equipment}
\newacronym{gcn}{GCN}{graph convolutional network}
\newacronym{spmm}{SpMM}{sparse-dense matrix multiply}
\newacronym{pnr}{P\&R}{place and route}
\newacronym{qor}{QoR}{quality of results}
\newacronym{ni}{NI}{network interface}
\newacronym{rob}{ROB}{reorder buffer}
\newacronym{lvds}{LVDS}{low-voltage differential signaling}
\newacronym{llm}{LLM}{large language model}
\newacronym{udp}{UDP}{user-defined primitives}
\newacronym{soc}{SoC}{system-on-chip}

\glsunset{cpu}
\glsunset{gpu}
\glsunset{dma}
\glsunset{simd}
\glsunset{ip}
\glsunset{fpu}
\glsunset{axi4}
\glsunset{soc}

\begin{document}

\title{Toward Open-Source Chiplets for HPC and AI:\\ {\socc} and Beyond 
}

\author{
    \IEEEauthorblockN{%
    Paul Scheffler\textsuperscript{\textasteriskcentered}, %
    Thomas Benz\textsuperscript{\textasteriskcentered}, %
    Tim Fischer\textsuperscript{\textasteriskcentered}, %
    Lorenzo Leone\textsuperscript{\textasteriskcentered}, %
    Sina Arjmandpour\textsuperscript{\textasteriskcentered}, %
    Luca Benini\textsuperscript{\textasteriskcentered}\textsuperscript{\textdagger}%
    }
    \IEEEauthorblockA{
        \textasteriskcentered~\textit{Integrated Systems Laboratory, ETH Zurich}, Switzerland \\
        \textdagger~\textit{Department of Electrical, Electronic, and Information Engineering, University of Bologna}, Italy \\
        \{paulsc,tbenz,fischeti,lleone,sarjmandpour,lbenini\}@iis.ee.ethz.ch
    }
}

\maketitle

\begin{abstract}
We present \lb{a} roadmap for open-source chiplet-based RISC-V systems targeting high-performance computing and artificial intelligence, aiming to close the performance gap to proprietary designs. 
Starting with \emph{\socc}, the first open, silicon-proven dual-chiplet RISC-V manycore in 12nm FinFET, we scale to \emph{\sflo}, a mesh-NoC-based dual-chiplet system, and to \emph{\sogo}, a 7nm quad-chiplet concept architecture achieving state-of-the-art compute density. 
Finally, we explore possible avenues to extend openness beyond logic-core RTL into simulation, EDA, PDKs, and off-die PHYs.
\end{abstract}


\section{Introduction}

%

%

%

Soaring compute demands and rapidly growing datasets in \gls{ai} and \gls{hpc} are outpacing the performance and memory bandwidth advances \lb{offered by technology scaling.}
\lb{As the gap widens with the slowdown of technology scaling,} system designers are turning to increasingly large and specialized architectures to sustain performance and energy efficiency needs. 
However, single-die packaging has emerged as a bottleneck, constraining designs in terms of area, yield, and connectivity~\cite{li2024highbandwidth}.

These limitations are driving a transition toward \emph{2.5D integration}, wherein multiple discrete chiplets are interconnected \lb{at short distance and fine pitch through an interposer.}
The yield, scale, and die-to-die bandwidth of 2.5D systems \lb{vastly exceed} those of single-die packages, unlocking further scaling.
Standardized interfaces and ecosystems, as proposed by the \emph{Open Chiplet Architecture}~\cite{tenstorrent2025oca}, would allow chiplets to serve as interchangeable \lb{building} blocks, enabling further architectural specialization and \lb{sustained} performance and efficiency gains.

\begin{figure}
\includegraphics[width=0.92\linewidth]{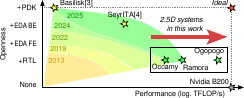}
\caption{Overview of the openness-performance tradeoff, exemplified by various open-source systems including those discussed in this paper.}
\label{fig:intro} 
\end{figure}

The idea of an open, composable chiplet ecosystem naturally aligns with the goals and principles of \emph{open-source hardware}, which has made significant progress in recent years~\cite{sauter2025basilisk}.
Like open-source \glspl{ip}, \emph{open-source chiplets} could become important building blocks in future 2.5D systems, providing many of the same benefits including high design transparency, low integration cost, and low barriers to design collaboration~\cite{li2024highbandwidth}. 
The main drawback of current open-source designs is their limited \emph{performance}:
as shown in \cref{fig:intro}, there is a significant performance gap between \gls{soa} and open-source systems demonstrated in silicon, which grows larger as the degree of openness increases.
\lb{For instance, systems designed with open \gls{eda} like \emph{SeyrITA}~\cite{bertuletti2025opensource} feature at most a few processing elements, while \emph{Basilisk}~\cite{sauter2025basilisk}, an \emph{end-to-end} open-source design using an open \gls{pdk}, features only a single core.} 
To enable the proliferation of open-source chiplets and their benefits, designers must close this performance gap by working toward open chiplet designs with competitive performance.

In this paper, we present our evolving roadmap on open-source 2.5D RISC-V manycores targeting \gls{hpc} and \gls{ai} workloads, addressing the need for high-performance open-source chiplet designs.
We discuss our \emph{past}, \emph{present}, and \emph{future} efforts on scaling the performance of open-source chiplet designs while prioritizing high compute utilization, energy efficiency, and compute density.

We begin with \emph{\socc}~\cite{scheffler2025occamy}, our first chiplet-based system and the first open-source 2.5D RISC-V manycore demonstrated in silicon.
\socc's 432 compute cores feature multiple \gls{isa} extensions to efficiently handle the mixed-regularity compute of modern \gls{hpc} and \gls{ai} workloads, 
achieving
a peak performance of \SI{876}{DP\dash\giga\flop\per\second} as well as competitive (\SI{89}{\percent}) and leading (42--\SI{83}{\percent}) peak \gls{fpu} utilizations on regular \lb{(dense)} and irregular \lb{(sparse)} workloads, respectively.
While {\socc}'s compute architecture proved highly effective, we \lb{found that} its interconnect, based on centralized crossbars, limits the scalability of its architecture to larger systems.

To address this, we \lb{propose} \emph{\sflo}~\cite{floonoc}, our second-generation dual-chiplet system featuring a scalable, low-overhead 2D mesh \gls{noc}.
{\sflo} extends \socc's clusters with co-integrated \gls{noc} routers to enable their 2D tiling, significantly reducing interconnect overheads.
This enables \SI{33}{\percent} more clusters and an \SI{11}{\percent} higher peak clock on the same chiplet area, raising compute density by \SI{43}{\percent} and increasing peak performance to \SI{1.29}{DP\dash\tera\flop\per\second}.
It also provides enough bandwidth at its mesh edge for a \SI{16}{\x} faster \SI{1.04}{\tera\bit\per\second} \gls{d2d} interface.
Despite these benefits, we identify further opportunities to improve on {\sflo}'s architecture by expanding the compute mesh and accelerating data movement in the \gls{noc}.

These insights lead us to \emph{\sogo}, a concept architecture expanding \sflo's compute mesh to a quad-chiplet, quad-HBM3, \SI{10.3}{DP\dash\tera\flop\per\second} system in \SI{7}{\nano\meter} FinFET.
In addition to its \SI{8.0}{\x} higher peak performance and \SI{2.8}{\x} higher compute density, {\sogo} introduces three lightweight extensions accelerating common \gls{noc} communication patterns: 
packed irregular streams, in-stream \gls{dma} operations, and in-router collectives.
With Ogopogo’s node-normalized compute density exceeding that of Nvidia's \gls{soa} B200 \gls{gpu} by \SI{19}{\percent}, we shift our attention to increasing the \emph{openness} of our 2.5D designs:
in addition to the open \gls{rtl} description of our logic cores, we explore the possibilities of open simulation setups, open \gls{eda} flows, open \glspl{pdk}, and even open off-die \glspl{phy}.

To summarize, our contributions are as follows:

\begin{itemize}

\item%
We review and discuss \emph{\socc} (\cref{sec:socc}), our silicon-proven dual-chiplet, dual-HBM2E, \SI{876}{DP\dash\giga\flop\per\second} RISC-V manycore in \SI{12}{\nano\meter} FinFET designed to efficiently handle both dense and sparse \gls{hpc} and \gls{ai} workloads.
\lb{We discuss \socc's main limitation, namely} its  interconnect based on \lb{hierarchical} crossbars\lb{, which limits its efficiency and scalability to larger systems}.

\item%
We present \emph{\sflo} (\cref{sec:sflo}), our second-generation dual-chiplet system tiling \socc's proven clusters in a scalable, low-overhead 2D mesh \gls{noc}.
On the same chiplet area, \sflo~provides \SI{33}{\percent} more clusters and a \SI{16}{\x} faster \gls{d2d} link, reaching up to 
\SI{1.29}{DP\dash\tera\flop\per\second}.
While \sflo~is highly area-efficient, we identify further opportunities in expanding its compute mesh and accelerating inefficient data movement in its \gls{noc}.

\item%
We propose \emph{\sogo} (\cref{sec:sogo}), a  \SI{7}{\nano\meter} quad-chiplet, quad-HBM3 concept architecture scaling up \sflo's compute mesh to \SI{10.3}{DP\dash\tera\flop\per\second}
while introducing three lightweight data movement extensions: 
in-router collectives, in-stream \gls{dma} operations, and packed irregular streams.
Ogopogo's node-normalized compute density exceeds that of Nvidia's B200 \gls{gpu} by \SI{19}{\percent}.

\item%
We explore how to expand the openness of future chiplet-based architectures (\cref{sec:e2ec}).
In addition to open-RTL logic cores, we explore the possibilities of open simulation, open \gls{eda}, and even open \glspl{pdk} and off-die \glspl{phy}.

\end{itemize}

\section{\socc: A Dual-Chiplet Silicon Demonstrator}
\label{sec:socc}

\emph{\socc}~\cite{scheffler2025occamy} is our first chiplet-based system and the first open-source 2.5D RISC-V manycore demonstrated in silicon.
It features two \SI{12}{\nano\meter} FinFET compute chiplets on a passive \SI{65}{\nano\meter} interposer, each paired with \SI{16}{\gibi\byte} of HBM2E DRAM and connected through a fully digital \gls{d2d} link.
Its 432 RV32G compute cores feature multiple \gls{isa} extensions~\cite{zaruba2020snitch,scheffler2023sssr,bertaccini2024fpu,benz2024idma} to efficiently handle the sparse and mixed-precision compute of modern \gls{hpc} and \gls{ai} applications,  
achieving %
competitive (\SI{89}{\percent}) and leading (42--\SI{83}{\percent}) peak \gls{fpu} utilizations on \lb{dense and sparse} workloads, respectively.
The cores are organized into 48 \cut{shared-memory }\emph{clusters}, which communicate through a hierarchical crossbar interconnect and leverage explicit, \gls{dma}-based data movement.

\cref{sec:socc:arch} presents \socc's hierarchical architecture.
In \cref{sec:socc:impl}, we shortly discuss \socc's silicon implementation.
\cref{sec:socc:res} presents our silicon evaluation, focusing on the compute utilization and energy efficiency benefits of \socc's \gls{isa} extensions and cluster architecture.
Finally, \cref{sec:socc:disc} will compare {\socc} to existing CPUs, GPUs, and accelerators, and discuss the scalability challenges arising from its hierarchical crossbar interconnect.

\subsection{Architecture}
\label{sec:socc:arch}

\begin{figure}[t]
    \centering%
    \begin{subcaptionblock}{0.22\linewidth}
        \centering%
        \includegraphics[width=\linewidth]{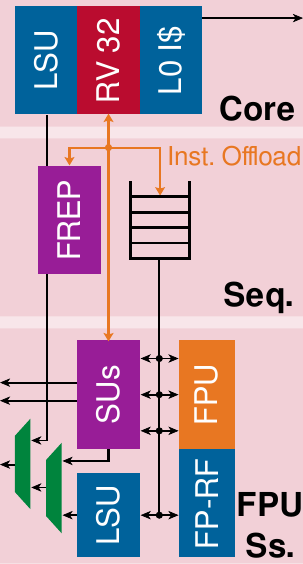}%
        \vspace{-1mm}
        \caption{Wk. core}%
        \vspace{2mm}
        \label{fig:socc-arch-core}%
    \end{subcaptionblock}\hfill
    \begin{subcaptionblock}{0.75\linewidth}
        \centering%
        \includegraphics[width=\linewidth]{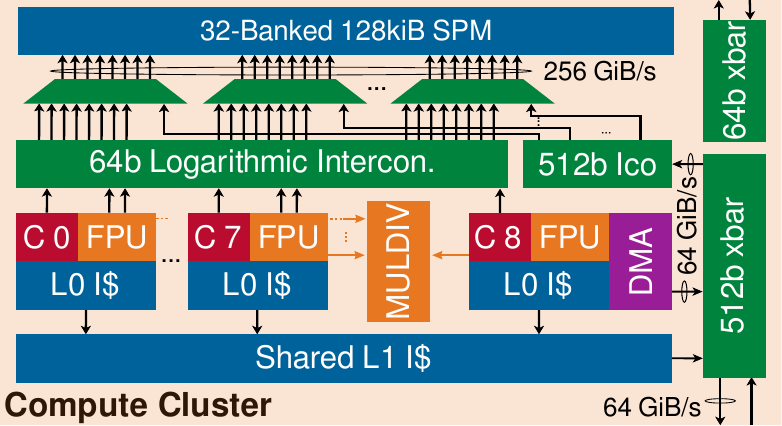}%
        \vspace{-1mm}
        \caption{Compute cluster}%
        \vspace{2mm}
        \label{fig:socc-arch-cluster}%
    \end{subcaptionblock}\hfill
    \begin{subcaptionblock}{\linewidth}
        \centering%
        \includegraphics[width=\linewidth]{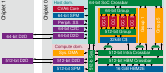}%
        \caption{Chiplet and system}%
        \label{fig:socc-arch-chiplet}%
    \end{subcaptionblock}
    \caption{Hierarchical architecture of {\socc}.}
    \label{fig:socc-arch}
\end{figure}

We present \socc's architecture in a bottom-up fashion.
\cref{fig:socc-arch-cluster} shows \socc's compute cluster, which is based on~\cite{zaruba2020snitch}.
Each cluster comprises eight worker cores and a \gls{dma} control core, which share a 32-bank, \SI{128}{\kibi\byte} \gls{spm} and an \SI{8}{\kibi\byte} L1 instruction cache.
Each worker core, shown in \cref{fig:socc-arch-core}, features a 64-bit-wide \gls{simd} \gls{fpu} with FP8-to-FP64 support.
In addition to standard multiply-accumulate instructions, the \gls{fpu} also supports widening sum-dot-product and three-addend summation for FP8 and FP16~\cite{bertaccini2024fpu}.

Two worker-core \gls{isa} extensions maximize \gls{fpu} utilization on both regular and irregular workloads: a floating-point hardware loop~\cite{zaruba2020snitch} and three sparsity-capable \glspl{su}~\cite{scheffler2023sssr}.
All three \glspl{su} support up to 4D strided accesses to the shared \gls{spm} to accelerate regular workloads. %
Two \glspl{su} additionally support indirect streams with 8-to-32-bit indices to accelerate scatter-gather accesses in \gls{spm}. 
Finally, both indirect \glspl{su} can cooperate to accelerate sparse tensor intersection and union, with the third \gls{su} optionally writing out the joint indices for sparse result tensors.
Together, each core's \glspl{su} enable a sustained bandwidth into the shared \gls{spm} of up to \SI{24}{\byte} per cycle.

The \gls{dma} control core features a tightly-coupled 512-bit \gls{dma} engine~\cite{benz2024idma}, which enables high-bandwidth. asynchronous $\leq$2D transfers between external memory %
and the cluster \gls{spm}.
This core coordinates the computation of worker cores and their fine-grained, low-latency accesses to the \gls{spm} with the latency-tolerant \gls{dma} transfers of large, double-buffered data tiles.
To reduce backpressure in the chiplet-level interconnect, the \gls{dma} engine has priority access to the shared \gls{spm} through a secondary interconnect;
it accesses contiguous blocks of eight banks at once, transferring up to \SI{64}{\byte} per cycle.

\cref{fig:socc-arch-chiplet} shows \socc's chiplet-level architecture, which comprises a \emph{compute} and a \emph{host} domain.
The compute domain is hierarchically organized into six \emph{groups} of four clusters each.
Grouped clusters %
have full-bandwidth access to each other through two \gls{axi4} crossbars:
a 512-bit crossbar used by \gls{dma} engines and instruction caches, and an atomics-capable 64-bit crossbar used for synchronization and message passing.
Each group shares one outgoing and one incoming port per crossbar to the chiplet-level interconnect, providing \SI{64}{\gibi\byte\per\second} for bulk transfers and \SI{8}{\gibi\byte\per\second} for message passing.
The outgoing 512-bit port additionally features a shared, configurable \SI{32}{\kibi\byte} constant cache.

At the chiplet level, the 512-bit ports of all six groups connect to an \emph{intra-group} crossbar, interconnecting the groups, and to an \emph{HBM} crossbar connecting to all eight HBM2E controllers.
A 512-bit \emph{system} crossbar bridges the 64-bit and 512-bit networks and provides access to a chiplet-level \gls{dma} engine, \SI{1}{\mebi\byte} of \gls{spm} for fast chiplet-level communication, as well as a 512-bit \gls{d2d} port.
All 64-bit group ports connect to the 64-bit \emph{SoC} crossbar, which also connects the host domain and a 64-bit D2D port.
The host domain is built around a Linux-capable RV64GC CVA6 core~\cite{zaruba2019cva6}, which is used to dispatch and orchestrate on-chiplet computation;
it provides various peripherals and a dedicated \SI{512}{\kibi\byte} \gls{spm} for host management tasks.
\cut{The peripherals include UART, I2C, QSPI, GPIOs, a \SI{1.33}{\giga\bit\per\second} off-interposer link, and a JTAG test access point for live host debugging.}
\cut{They also include RISC-V-compliant timers and platform-level interrupt controllers, providing interrupts for both the host processor and all on-chip compute cores.}

The \cut{inter-chiplet }\gls{d2d} link features two segments connecting to the two on-chip networks: 
a \emph{narrow} (64-bit) segment with a single \gls{phy} and a \emph{wide} (512-bit)  segment with 38 \glspl{phy}.
Both segments serialize their networks' \gls{axi4} transactions to \emph{AXI-Stream} for transmission\cut{, exerting backpressure on the sending interfaces to avoid deadlocks}.
Each \gls{phy} features an all-digital, source-synchronous duplex interface with eight \gls{ddr} lanes in each direction.
The transmitter side forwards a divided system clock, while the receiver side synchronizes packets to its system clock to reassemble them.
The wide segment features a \emph{channel allocator} to enable fault tolerance: an initial calibration detects faulty \glspl{phy}, which can individually be disabled.
The narrow and wide segments achieve effective duplex bandwidths of up to \SI{1.33}{\giga\bit\per\second} and \SI{64}{\giga\bit\per\second}, respectively.

\subsection{Implementation}
\label{sec:socc:impl}

\begin{figure}[t]
    \centering%
    \begin{subcaptionblock}{0.42\linewidth}
        \includegraphics[width=\linewidth]{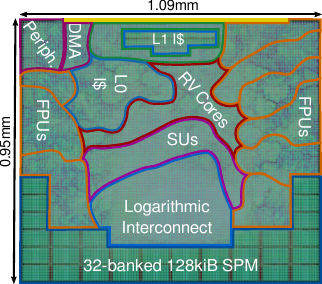}%
        \caption{Cluster layout}%
        \label{fig:socc-impl-cluster}%
        \includegraphics[width=\linewidth]{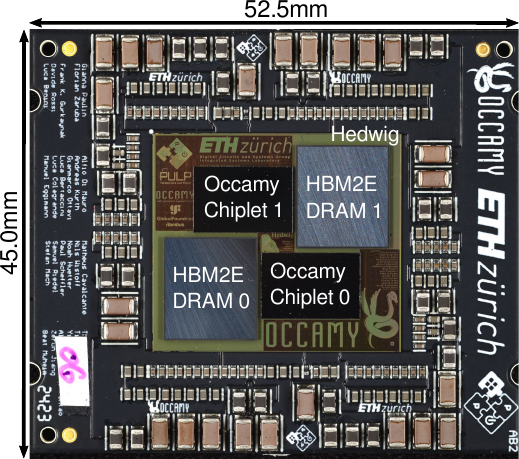}%
        \caption{Assembled module}%
        \label{fig:socc-impl-chiplet}%
    \end{subcaptionblock}
    \begin{subcaptionblock}{0.535\linewidth}
        \includegraphics[width=\linewidth]{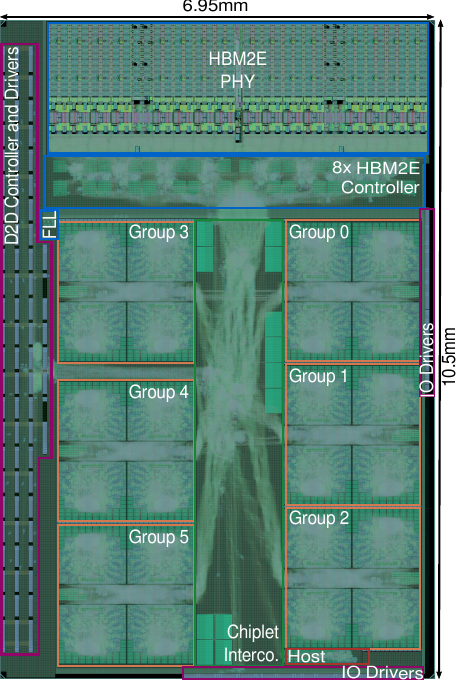}%
        \caption{Chiplet layout}%
        \label{fig:socc-impl-system}%
    \end{subcaptionblock}
    \caption{Physical implementation of {\socc}.}
    \label{fig:socc-impl}
\end{figure}

The full 2.5D \socc~system was implemented and fabricated along with a carrier board, resulting in the compute module shown in \Cref{fig:socc-impl-chiplet}.
The two compute dies with their respective Micron \emph{MT54A16G808A00AC-32} HBM2E stacks were mounted on \emph{Hedwig}, a passive, four-layer \SI{65}{\nano\meter} PKG-25SI interposer from \gf.

The compute chiplets were implemented in \gf' \SI{12}{\nano\meter} LP+ FinFET technology using Synopsys \emph{Fusion Compiler}. 
We used 7.5-track standard cells from Arm and HBM2E \glspl{ip} (controller and \gls{phy}) provided by Rambus.
We targeted a nominal compute clock of \SI{1}{\giga\hertz}, achieving up to \SI{1.14}{\giga\hertz} under typical conditions (\SI{0.8}{\volt}, \SI{25}{\celsius}), and \SI{0.95}{\giga\hertz} under worst-case conditions (\SI{0.72}{\volt}, \SI{125}{\celsius}).

\cref{fig:socc-impl-cluster,fig:socc-impl-chiplet} show the annotated physical layouts of the compute cluster and chiplet, respectively.
The \SI{1.0}{\milli\meter^2} cluster is area-dominated by its nine compute cores (\SI{44}{\percent}) and \gls{spm} (\SI{17}{\percent}).
Each group (G$n$) is dominated by its clusters (\SI{83}{\percent}).
The top-level chiplet layout is dominated by its compute groups (\SI{39}{\percent}), HBM2E interface (\SI{25}{\percent}), and \gls{d2d} link (\SI{11}{\percent}).
\cut{In accordance with the interposer arrangement,} 
The west and south chiplet edges are reserved for the HBM2E and \gls{d2d} interfaces, respectively, while off-interposer IO drivers are kept on the north and west edges.

\subsection{Evaluation}
\label{sec:socc:res}

We evaluated \socc's performance, power consumption, and energy efficiency on various dense, sparse, and mixed-regularity workloads with and without \gls{su} acceleration. 
We provided its nominal \SI{1}{\giga\hertz} clock and typical operating conditions (\SI{25}{\celsius}, \SI{0.8}{\volt}).

\Cref{fig:socc-res-res64} presents our results on increasingly irregular double-buffered FP64 workloads.
\socc's \glspl{su} accelerate dense \gls{gemm} by \SI{2.7}{\x}, reaching a near-ideal \gls{fpu} utilization of \SI{89}{\percent}, a performance of \SI{686}{\giga\flop\per\second}, and an energy efficiency of \SI{39.8}{\giga\flop\per\second\per\watt}.
Stencil computations (STC) are accelerated by up to \SI{3.9}{\x}, reaching up to \SI{83}{\percent} \gls{fpu} utilization, \SI{571}{\giga\flop\per\second}, and \SI{28.1}{\giga\flop\per\second\per\watt}. 
On \gls{gcn} layers, %
\socc's \glspl{su} provide speedups of up to \SI{2.3}{\x} and reach up to \SI{54}{\percent} \gls{fpu} utilization, \SI{413}{\giga\flop\per\second}, and \SI{25.0}{\giga\flop\per\second\per\watt}. 
Finally, \glspl{su} accelerate \gls{spmm} by up to \SI{4.6}{\x}, achieving up to \SI{42}{\percent} \gls{fpu} utilization, \SI{307}{\giga\flop\per\second}, and \SI{16.0}{\giga\flop\per\second\per\watt}.

\Cref{fig:socc-res-mpgemm} shows \socc's performance and energy efficiency on \gls{su}-accelerated \gls{gemm} with decreasing precision. 
We consider FP16 \gls{gemm} with and without expanding FP32 accumulation (EXP suffix) and FP8 \gls{gemm} only with expanding FP16 accumulation.
As we reduce precision, the performance and energy efficiency scale as expected, incurring slight conversion and packing overheads.
On expanding FP8 \gls{gemm}, \socc~achieves a throughput of \SI{4.1}{QP\text{-}\tera\flop\per\second} and an energy efficiency of \SI{263}{QP\text{-}\giga\flop\per\second\per\watt}.
On FP16, expanding \gls{gemm} is {\SI{6.5}{\percent}~more energy-efficient than non-expanding \gls{gemm} thanks to dedicated expanding dot product units in our \gls{fpu}~\cite{bertaccini2024fpu}.

\begin{figure}[t]
    \centering%
    \begin{subcaptionblock}{\linewidth}
        \centering%
        \includegraphics[width=\linewidth]{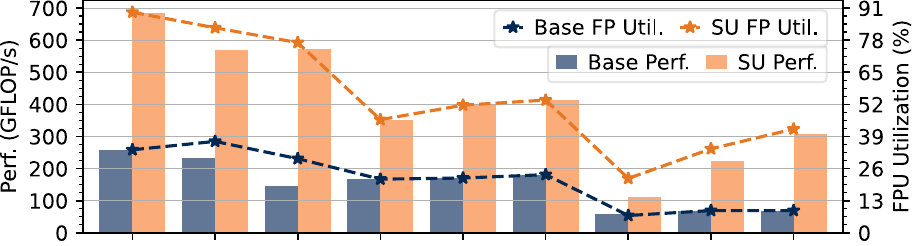}\\
        \vspace{0.5em}
        \hspace{0.015\linewidth}\includegraphics[width=0.985\linewidth]{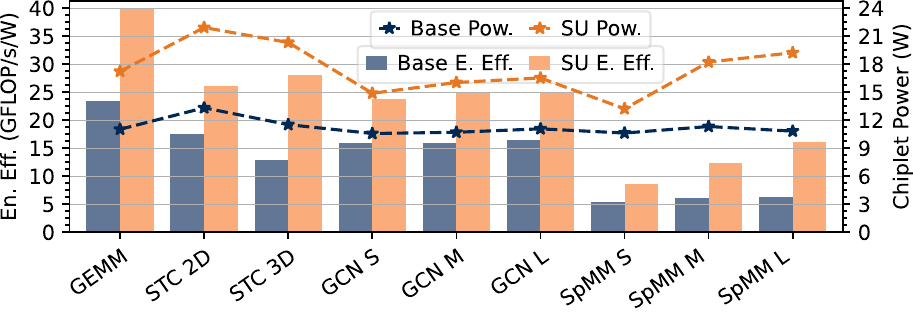}%
        \vspace{-0.3em}
        \caption{Dense and sparse double-buffered FP64 workloads}%
        \vspace{0.3em}
        \label{fig:socc-res-res64}%
    \end{subcaptionblock}

    \begin{subcaptionblock}{0.47\linewidth}
        \centering%
        \vspace{0.7em}
        \includegraphics[width=\linewidth]{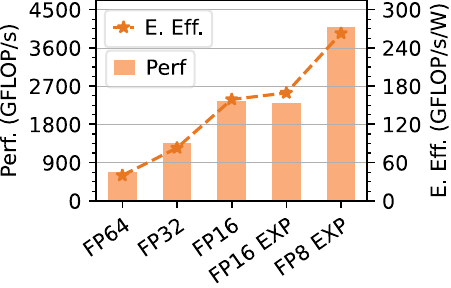}%
        \vspace{-0.35em}
        \caption{GEMM with decreasing \\precisions from FP64 to FP8}%
        \label{fig:socc-res-mpgemm}
    \end{subcaptionblock}%
    \hspace{0.5em}
    \begin{subcaptionblock}{0.47\linewidth}
        \centering%
        \vspace{0.7em}
        \includegraphics[width=\linewidth]{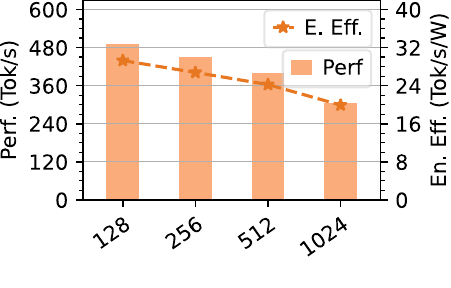}%
        \vspace{-0.4em}
        \caption{FP16 GPT-J LLM inference for increasing sequence lengths}%
        \label{fig:socc-res-llm}
    \end{subcaptionblock}
    
    \caption{Performance, power, and energy efficiency results for {\socc}. %
    }
    \label{fig:socc-res}
\end{figure}

\Cref{fig:socc-res-llm} presents \socc's performance and energy efficiency on \gls{su}-accelerated FP16 GPT-J inference in non-autoregressive mode.
We leverage the implementation presented in~\cite{potocnik2024optimizing}, 
using \emph{FlashAttention-2} for the attention mechanism\cut{ and fusing it with the subsequent concatenation and linear projection}.
At the shortest considered sequence length of 128, \socc~achieves a peak throughput of \SI{490}{\tok\per\second} and an energy efficiency of \SI{29.3}{\tok\per\second\per\watt}.
Increasing the sequence length gradually decreases throughput
due to the quadratic scaling of the attention computation.
Accordingly, we also see a gradual decrease in energy efficiency,  %
as less time is spent on matrix multiplication and more time on the less efficient element-wise \emph{softmax} operation.

Finally, we evaluate the utilization, energy efficiency, and latency of our fully digital \gls{d2d} link. 
On a \SI{16}{\kibi\byte} transfer, the \SI{64}{\giga\bit\per\second} wide segment achieves a link utilization of  \SI{96}{\percent} and an energy efficiency of \SI{1.6}{\pico\joule\per\bit}.
Using the host core to access the other chiplet's host \gls{spm} through the narrow segment incurs 27 cycles of latency, while a \gls{dma} transfer over the wide segment exhibits 61 cycles of latency.

\subsection{SoA Comparison and Discussion}
\label{sec:socc:disc}

We extensively compared {\socc} to \gls{soa} CPUs, GPUs, and accelerators~\cite{scheffler2025occamy}.
Unlike domain-specific accelerators targeting sparse compute, {\socc} is a \emph{general-purpose} system that efficiently handles both one- and two-sided sparse operations across a wide sparsity range and with high dataflow flexibility.
Its peak performance far exceeds most academic RISC-V prototypes and is comparable to commercial products like Sophon's SG2042~\cite{sophon2023trm}.
On dense workloads, its \SI{89}{\percent} \gls{fpu} utilization is competitive with \gls{soa} CPUs and GPUs, and its energy efficiency roughly matches that of Nvidia's A100~\cite{strohmaier2024top500}.
On stencil codes, \socc's \glspl{su} enable a leading \gls{fpu} utilization and node-normalized area efficiency, outperforming A100 by \SI{1.7}{\x}~\cite{zhang2023revisiting} and \SI{1.2}{\x}~\cite{chen2024convstencil}, respectively.
On sparse-dense linear algebra, {\socc} further expands its lead on \gls{soa} CPUs and GPUs, achieving a \SI{5.2}{\x}~\cite{alappat2024level} higher \gls{fpu} utilization and an \SI{11}{\x}~\cite{alappat2020a64fx} higher node-normalized area efficiency than leading competitors.
To summarize, {\socc}'s \gls{fpu} utilization and efficiency are \emph{competitive} on regular workloads while \emph{leading} the \gls{soa} on irregular workloads.

While {\socc}'s cluster-based compute architecture is highly effective across various workloads, its hierarchical crossbar-based interconnect %
poses significant scalability challenges for future systems.
Despite our careful hierarchical design, \SI{31}{\percent} of the compute domain's area is dedicated to crossbars, leaving only \SI{65}{\percent} to the compute clusters.
As we scale up {\socc}'s architecture, several fundamental limitations of this interconnect emerge.
First, crossbar area scales with the \emph{product} of requestor and endpoint counts, progressively reducing the area fraction available for compute.
Second, the centralized architecture of crossbars leads to longer global routes and higher routing congestion, eventually degrading the \gls{qor} despite parameter tuning and numerous routing iterations.
Finally, crossbar-based interconnects limit off-die bandwidth, as routing traffic from any requestor to any off-die channel requires increasingly large crossbars.
In fact, {\socc}'s HBM crossbar required careful design and tuning to maintain both physical implementability and full HBM bandwidth.
While this was manageable with {\socc}'s low-throughput, fully digital \gls{d2d} link, adopting an \gls{soa} high-bandwidth \gls{d2d} interface alongside HBM2E would have significantly complicated implementation.

To overcome these overheads and scalability limitations, a common solution is to adopt a mesh-based \gls{noc}~\cite{mckeown2018power, esp_isscc24, floonoc}.
Unlike crossbars that funnel traffic through a few congested nodes, a \gls{noc} distributes communication across routers and links that scale \emph{linearly}, rather than quadratically, with system size.
\cut{Through wide physical links, mesh \glspl{noc} can sustain high bandwidth to both on-chip endpoints and to off-chip links.}
By integrating routers into compute tiles, connections can be formed through abutment, \cut{greatly }simplifying implementation.
In the next section, we therefore present an {\socc}-\lb{scale} system that tiles its \cut{compute }clusters into a scalable mesh \gls{noc} to overcome the design, area, and implementation challenges we identified above.

\section{\sflo: A Scalable Cross-Chiplet Mesh}
\label{sec:sflo}

\emph{\sflo}~\cite{floonoc} is our second-generation open-source 2.5D system\lb{.}
It maintains \socc's highly effective cluster architecture%
, but adopts a \cut{wide-link }2D mesh \gls{noc} to reduce interconnect area overheads and improve system scalability.
Compared to \socc, {\sflo} features \SI{33}{\percent} more clusters on the same chiplet area and reaches an \SI{11}{\percent} higher clock speed, achieving a \SI{43}{\percent} higher compute density and a peak performance of \SI{1.29}{\tera\flop\per\second}.
Its \SI{0.15}{\pico\joule\per\byte\per hop} \gls{noc} increases HBM2E bandwidth utilization in low- and high-traffic scenarios by \SI{11}{\percent} and \SI{22}{\percent}, respectively,
and can accommodate an improved four-port \gls{d2d} link architecture with a \SI{16}{\x} higher wide-data bandwidth of \SI{1.04}{\tera\bit\per\second}.

In \cref{sec:sflo:arch}, we present \sflo's architectural innovations over {\socc}, focusing on its mesh \gls{noc} and improved \gls{d2d} link. 
\cref{sec:sflo:impl} discusses the implementation of {\sflo}'s compute domain and a possible chiplet implementation.
\cref{sec:sflo:res} presents our evaluation, focusing on the bandwidth, latency, and performance improvements enabled by {\sflo}'s \gls{noc}. 
Finally, \cref{sec:sflo:disc} will compare {\sflo} to existing mesh \gls{noc} systems and discuss possible next steps to scale up and improve our chiplet architecture.

\subsection{Architecture}
\label{sec:sflo:arch}

{\sflo} uses \cut{the open-source }\emph{FlooNoC}~\cite{floonoc} as its top-level interconnect, a wide-link mesh \gls{noc} serializing both narrow (64-bit) and wide (512-bit) \gls{axi4} transactions.
Each link consists of three statically routed physical channels:
the unidirectional \emph{req} and \emph{rsp} channels transport requests and 64-bit data, while the bidirectional \emph{wide} channel transports 512-bit bulk data.
FlooNoC provides two \gls{ni} variants, one with a \gls{rob} and one without.
The \gls{rob}-less \gls{ni} stalls outgoing requests whenever necessary to enforce ordering, \lb{requiring an out-of-order-capable requestor to maintain high throughput.}

{\sflo} largely reuses {\socc}'s cluster design, but extends it with \lb{an integrated} \gls{ni} and router so it can be tiled into a mesh.
The resulting architecture is shown in \cref{fig:sflo-arch-cluster}.
The cluster's \gls{ni} is \emph{\gls{rob}-less}: instead of incurring a large \gls{rob}\cut{ in each cluster}, the \gls{dma} engine is extended to distribute transfers onto multiple backends\cut{with different IDs}~\cite{benz2024idma}, enabling out-of-order processing. %
The cluster router fully interconnects five ports, one for each cardinal direction and one for the local \gls{ni}.
It supports multiple static routing algorithms (source-based, dimension-ordered, or table-based) as well as wormhole routing to prevent write burst interleaving.
Instead of leveraging virtual channels, each of the three physical channels is routed independently, ensuring network-level isolation.

\begin{figure}[t]
    \centering%
    \begin{subcaptionblock}{\linewidth}
        \centering%
        \includegraphics[width=\linewidth]{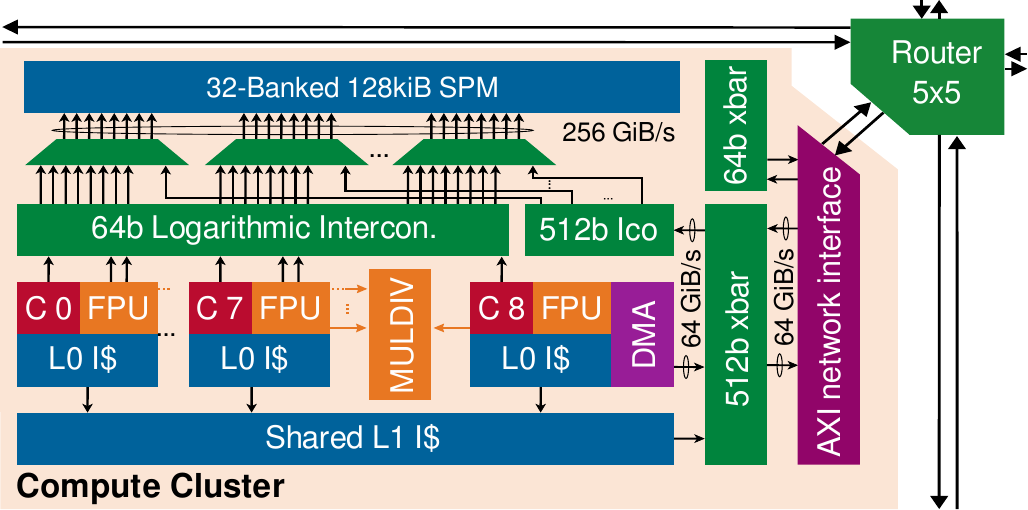}%
        \vspace{-1mm}
        \caption{Tilable cluster with its integrated \gls{ni} and router}%
        \vspace{2mm}
        \label{fig:sflo-arch-cluster}%
    \end{subcaptionblock}\hfill
    \begin{subcaptionblock}{\linewidth}
        \centering%
        \includegraphics[width=\linewidth]{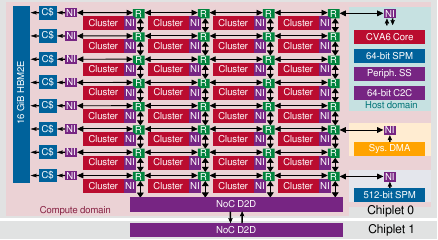}%
        \vspace{-1mm}
        \caption{Chiplet and system}%
        \vspace{-1mm}
        \label{fig:sflo-arch-chiplet}%
    \end{subcaptionblock}
    \caption{Hierarchical architecture of {\sflo}.}
    \label{fig:sflo-arch}
\end{figure}

\cref{fig:sflo-arch-chiplet} shows \sflo's compute chiplet architecture. 
As we will show in \cref{sec:sflo:impl}, the \gls{noc}'s significantly lower area overhead allows us to increase the per-chiplet cluster count from \cut{{\socc}'s }24 to 32 \emph{without} increasing chiplet area.
We arrange the clusters in an 8$\times$4 mesh.
On the left edge, we connect {\socc}'s existing HBM2E interface:
we connect each of the eight HBM2E controllers through their own \glspl{ni} and \SI{32}{\kibi\byte} constant caches, which were previously part of the compute groups.
On the right edge, we connect the existing host domain, the \SI{1}{\mebi\byte} system \gls{spm}, and the system-level \gls{dma} engine through their own \glspl{ni}.
On the bottom edge, we connect an improved \gls{d2d} interface to another chiplet through all four available \gls{noc} links.

The increased edge bandwidth provided by the mesh \gls{noc} enables the use of a significantly higher-throughput \gls{d2d} interface.
We replace \socc's bit-parallel \gls{d2d} \glspl{phy}, which use a divided system clock, with serial \SI{4}{\giga\bit\per\second} \gls{lvds} \glspl{phy} forwarding their own clock.
We match \socc's existing \gls{d2d} macro area of \SI{9.6}{\milli\meter^2}, providing 96, 20, and 18 \gls{phy} pairs for the \emph{wide}, \emph{req}, and \emph{rsp} channels of each \gls{noc} link, respectively.
We adapt the existing digital \gls{phy} frontend to directly transmit \gls{noc} flits, forming a virtual 16$\times$4 mesh that spans across both chiplets.
In total, \sflo's \gls{d2d} link provides an effective duplex bandwidth of \SI{1.04}{\tera\bit\per\second} for wide transfers and \SI{137}{\giga\bit\per\second} for narrow transfers.

\subsection{Implementation}
\label{sec:sflo:impl}

\begin{figure}[t]
    \centering%
    \begin{subcaptionblock}{0.40\linewidth}
        \centering%
        \includegraphics[width=\linewidth]{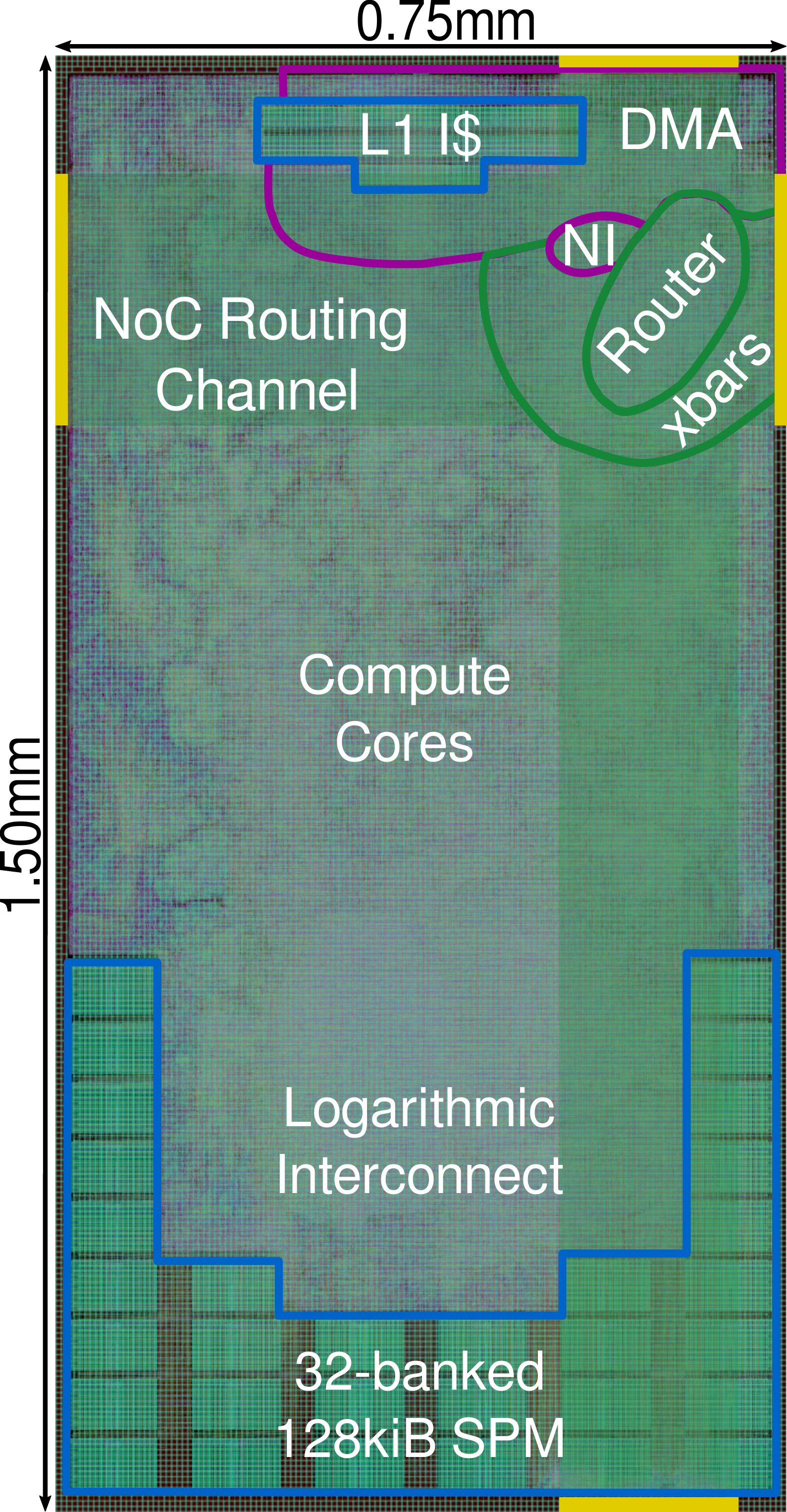}%
        \caption{Cluster tile layout}%
        \vspace{0mm}
        \label{fig:sflo-impl-cluster}%
    \end{subcaptionblock}\hfill
    \begin{subcaptionblock}{0.582\linewidth}
        \centering%
        \includegraphics[width=\linewidth]{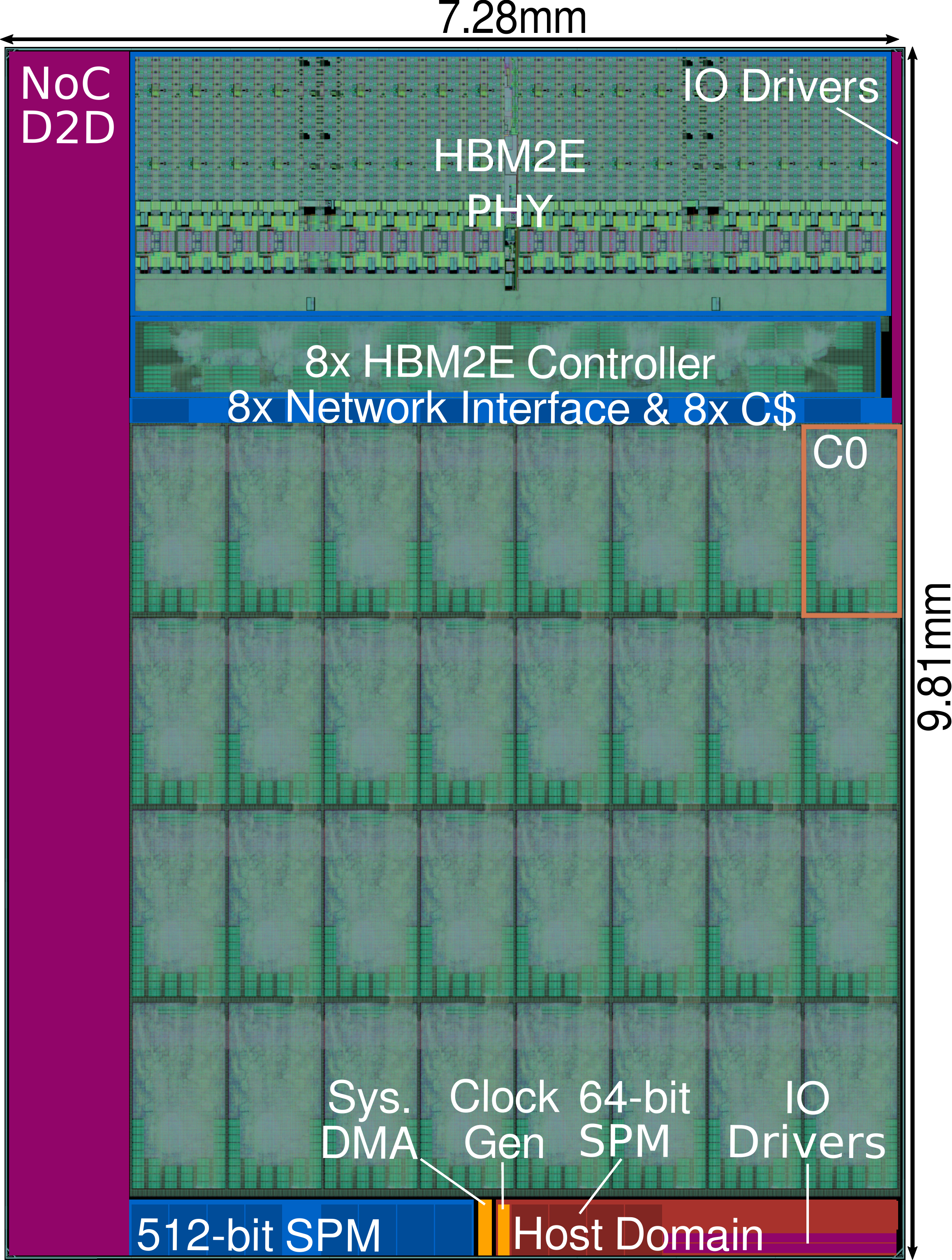}%
        \caption{Comp. dom. \& possible chiplet layout}%
        \label{fig:sflo-impl-chiplet}%
    \end{subcaptionblock}
    \caption{Physical implementation of {\sflo}.}
    \label{fig:sflo-impl}
\end{figure}

We implement {\sflo}'s clusters and compute domain in \gf's \SI{12}{\nano\meter} LP+ FinFET technology and estimate a full chiplet implementation based on them.
We use Synopsys \emph{Fusion Compiler} for synthesis, placement, and routing with Arm 7.5-track standard cells, and reuse \socc's HBM2E interface.
As for \socc, we target a nominal compute clock of \SI{1}{\giga\hertz}, achieving up to \SI{1.26}{\giga\hertz} under typical conditions and \SI{0.85}{\giga\hertz} under worst-case conditions.

\cref{fig:sflo-impl-cluster} shows \sflo's cluster layout.
To approximately match the per-port HBM2E interface height, we reshape the cluster to a 2:1 aspect ratio.
We increase the cluster area by \SI{8.6}{\percent} to accommodate the router, the \gls{ni}, and our multi-backend \gls{dma} engine extensions.
The \gls{noc} links are routed on the upper metal layers, leveraging free routing resources above existing logic and memory macros.

\cref{fig:sflo-impl-chiplet} shows \sflo's implemented compute domain and its integration into a possible chiplet implementation.
To allow for clock tree and global control signal routing, we leave small gaps between the cluster tiles, bridging \gls{noc} links across them.
We estimate and allocate the area of remaining blocks including the host domain, system \gls{spm}, HBM2E glue logic, and off-interposer IO as shown. 
We further maintain the \gls{d2d} link's existing \SI{9.6}{\milli\meter^2} area as explained in \cref{sec:sflo:arch}.
The resulting \SI{71}{\milli\meter\squared} chiplet is almost the exact same size as {\socc}'s (\SI{2.1}{\percent} smaller) despite featuring \SI{33}{\percent} more compute and a \SI{16}{\x} higher wide-transfer \gls{d2d} bandwidth.

\subsection{Evaluation}
\label{sec:sflo:res}

We evaluate {\sflo}'s on-chip latency, HBM bandwidth utilization, and peak performance and compare them to those of \socc.
We focus on evaluating the benefits of \sflo's \gls{noc}, as both systems share the same compute cluster architecture.

\cref{fig:sflo-res-lat} compares the average, minimum, and maximum inter-cluster latency under zero and full interconnect load. 
Overall, \sflo's \gls{noc} trades \lb{a modest} latency \lb{increase} for \lb{a} significantly improved scalability,
as many lightweight routers and links replace a few \lb{huge} crossbars.
Minimum latencies are higher on {\sflo} due to \socc's dedicated inter-group links,
whereas maximum latencies are higher due to the linear increase of hop counts in 2D meshes.
Average latencies, however, are very similar in both systems, being \SI{0.4}{\percent} smaller in {\sflo} under zero load and only \SI{8.4}{\percent} higher under full load.
Thus, {\sflo} manages to \emph{combine} the low average latencies of crossbar-based interconnects with the high scalability of a 2D mesh.

\cref{fig:sflo-res-bw} compares the single-channel HBM2E bandwidth utilization of one cluster's \gls{dma} engine under zero and full HBM load from other clusters.
The peak achievable full-load utilization is \SI{25}{\percent}, as four clusters each share the bandwidth of one 512-bit link.
Under zero load, {\sflo} notably increases the average bandwidth utilization from \SI{84.7}{\percent} to \SI{96.6}{\percent}, thanks to fewer hops and routing steps to the memory controllers.
Under full load, {\sflo}'s advantage becomes even more pronounced, increasing average utilization from \SI{20.2}{\percent} to \SI{24.8}{\percent}.
This is because HBM channels are no longer accessed through a single crossbar, but distributed among edge routers, reducing contention.
Overall, \sflo's decentralized mesh routing enables significant HBM bandwidth utilization benefits.

\begin{figure}[t]
    \centering%
    \begin{subcaptionblock}{0.3\linewidth}
        \centering%
        \includegraphics[width=\linewidth]{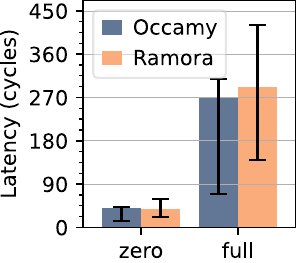}%
        \caption{On-chip latency}%
        \label{fig:sflo-res-lat}%
    \end{subcaptionblock}\hfill
    \begin{subcaptionblock}{0.3\linewidth}
        \centering%
        \includegraphics[width=\linewidth]{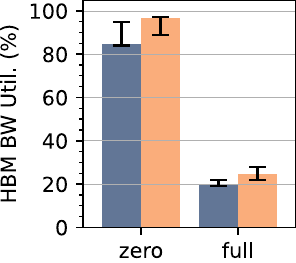}%
        \caption{HBM bw. util.}%
        \label{fig:sflo-res-bw}%
    \end{subcaptionblock}\hfill
    \begin{subcaptionblock}{0.3\linewidth}
        \centering%
        \includegraphics[width=\linewidth]{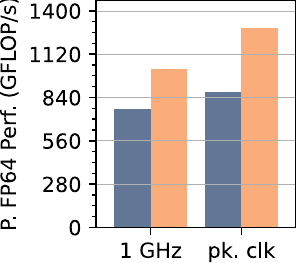}%
        \caption{Peak FP64 perf.}%
        \label{fig:sflo-res-perf}%
    \end{subcaptionblock}
    \caption{Inter-cluster latency, HBM bandwidth utilization, and peak FP64 performance of {\sflo} compared to \socc. Error bars indicate the full result range across clusters (minimum to maximum).}
    \label{fig:sflo-res}
\end{figure}

\cref{fig:sflo-res-perf} compares the peak FP64 performance of {\sflo} and {\socc}, both at \SI{1}{\giga\hertz} and at their respective \SI{1.26}{\giga\hertz} and \SI{1.14}{\giga\hertz} peak clocks under typical conditions.
At \SI{1}{\giga\hertz}, {\sflo}'s additional clusters increase peak performance from 768 to \SI{1024}{\giga\flop\per\second}.
At peak clock, {\sflo}'s further expands its lead thanks to its \SI{11}{\percent} higher clock speed, increasing peak performance from 876 to \SI{1290}{\giga\flop\per\second}.
Thus, {\sflo} significantly improves on {\socc}'s peak performance by increasing both its cluster count and its peak clock speed.

Finally, we assess the energy efficiency of \sflo's \gls{noc}. 
We estimate the power consumption of an otherwise idle cluster performing a \SI{4}{\kibi\byte} \gls{dma} transfer to a neighboring tile in post-layout simulation.
Of the cluster tile's \SI{128}{\milli\watt} draw, only \SI{15}{\percent} were consumed by the components actively involved in the transfer.
The routers themselves consumed only \SI{596}{\pico\joule}, yielding an energy efficiency of \SI{0.15}{\pico\joule\per\byte\per hop}.

\subsection{SoA Comparison and Discussion}
\label{sec:sflo:disc}

We compared {\sflo} to \gls{soa} 2D mesh \gls{noc} systems~\cite{floonoc} including \emph{Piton}~\cite{mckeown2018power}, \emph{Celerity}~\cite{rovinski2019evalcel}, and \emph{ESP}~\cite{esp_isscc24}.
Among all considered systems, {\sflo} achieves the highest simplex tile-to-tile bandwidth and energy efficiency, \SI{2.6}{\x} and \SI{3.0}{\x} higher than the runner-ups ESP and Celerity, respectively.
At the same time, the \gls{noc}'s power contribution in {\sflo} is \SI{4.0}{\x} lower than in ESP.
{\sflo} has the second-lowest \gls{noc} area overhead after Piton, which provides much lower bandwidths and narrower links.
Only ESP was evaluated in an equally advanced technology (also \SI{12}{\nano\meter} FinFET), while only Celerity can exceed \sflo's peak clock speed.
While Celerity's larger 8$\times$62 mesh naturally provides a larger aggregate NoC bandwidth,
{\sflo} features the largest total link data width.
Overall, while {\sflo} does not feature the largest 2D mesh \gls{noc} yet, its wide, low-overhead links enable leading tile-to-tile bandwidth and energy efficiency.

With \sflo, we combined \socc's proven compute cluster architecture with a scalable, low-overhead 2D mesh \gls{noc} providing leading throughput and efficiency.
However, there are still opportunities to improve our open-source chiplet-based designs.
First and foremost, {\sflo}'s highly scalable \gls{noc} allows for even larger meshes:
by increasing the chiplet size and moving to a more advanced technology node, we could further expand the compute domain.
This would also provide more edge links, enabling even higher HBM and \gls{d2d} bandwidths;
we could even leverage the remaining free mesh edge to add a second \gls{d2d} link, enabling quad-chiplet systems.
Secondly, we could extend the clusters, routers, and endpoints with specialized hardware units to reduce \gls{noc} communication overheads.
By reducing redundant and inefficient \gls{noc} traffic, we can reduce the energy cost of both on- and off-chip data movement and accelerate memory-bound workloads, improving both performance and energy efficiency.

Based on these insights, the next section will present the concept architecture of our next open-source 2.5D design, which scales \sflo's architecture to four larger chiplets while introducing three new extensions accelerating common \gls{noc} communication patterns.

\section{\sogo: A Quad-Chiplet Concept Architecture}
\label{sec:sogo}

\emph{\sogo} is a quad-chiplet \SI{7}{\nano\meter} concept architecture that scales up {\sflo}'s compute mesh to an \SI{8}{\x} higher peak performance.
It introduces three lightweight extensions improving data movement efficiency: in-router collectives, in-stream \gls{dma} operations, and packed irregular streams.
To sustain the higher edge bandwidth of its larger chiplet mesh, \sogo~upgrades from HBM2E to HBM3, doubling the per-chiplet bandwidth.
An estimated implementation of Ogopogo achieves a peak performance of \SI{10.3}{DP\dash\tera\flop\per\second} and a peak compute density of \SI{41.1}{DP\dash\giga\flop\per\second\per\milli\meter^2}, \SI{19}{\percent} higher than that achieved by Nvidia's B200 \gls{gpu}~\cite{nvidia2024blackwell} when normalized to our \SI{7}{\nano\meter} target node.

In \cref{sec:sogo:arch}, we will describe \sogo's quad-chiplet architectures and its data movement extensions.
In \cref{sec:sogo:eval}, we will evaluate a possible implementation of \sogo~and compare the full evolution of our 2.5D manycore designs to Nvidia's \gls{soa} B200 \gls{gpu}.
In the following \cref{sec:e2ec}, we will explore avenues to expand the openness of our 2.5D systems beyond logic-core \gls{rtl} descriptions.

\subsection{Architecture}
\label{sec:sogo:arch}

\Cref{fig:sogo-arch-top} shows \sogo's concept architecture.
Each of the four \SI{7}{\nano\meter} chiplets features a 16$\times$8 compute mesh, reusing \sflo's scalable \gls{noc} architecture and introducing the data movement extensions we describe below.
The left mesh edge now provides a \SI{6.4}{\giga\bit\per\second\per pin} HBM3 interface, whose increased bandwidth and parallelism can serve requests from all 16 edge routers.
The bottom and right mesh edges feature \gls{d2d} links reusing \sflo's \gls{lvds} \glspl{phy}, which provide wide-transfer bandwidths of \SI{1.02}{\tera\bit\per\second}} and \SI{2.05}{\tera\bit\per\second} to the respective neighboring chiplets.
The top mesh edge now connects the remaining chiplet resources including the host domain, system \gls{dma}, system \gls{spm}, and off-interposer IO.
We update the host domain to our standardized \emph{Cheshire} platform~\cite{ottaviano2023cheshire}, which supports multi-core hosts, multi-port \gls{noc} connections, and additional peripherals.

\begin{figure}[t]
    \centering%
        \centering%
        \includegraphics[width=\linewidth]{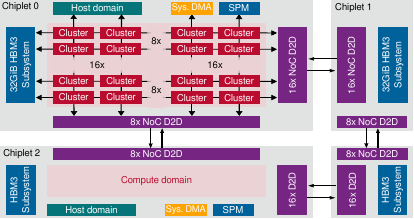}%
        \vspace{0.3em}
        \caption{{\sogo} chiplet and system architecture.}%
        \label{fig:sogo-arch-top}%
    \label{fig:sogo-arch}
\end{figure}

\sogo's \gls{dma} engines, routers, and HBM3 endpoints feature three lightweight hardware extensions which improve \gls{noc} bandwidth efficiency and accelerate common communication patterns:

\paragraph*{In-router collectives} We extend the routers with \emph{fork} and \emph{join} primitives to handle collective communication patterns like \emph{multicast}, \emph{broadcast}, and \emph{barriers} inside the \gls{noc} rather than at the endpoints. 
To implement multicast and broadcast, requests are forked (replicated) along their path to multiple destinations, and the corresponding responses are joined (reduced) as they return to the source.
To implement barrier synchronization, we instead join the inflight requests from multiple sources and then fork the response back to all sources. 
These capabilities come at a modest cost, increasing router area by less than $10$\% without degrading system timing.

\paragraph*{In-stream operations} We extend the cluster \gls{dma} engines with \emph{in-stream} vector units, which can perform simple element-wise and reductive operations on the fly.
Once read in by the engine, the data stream is aligned to its element boundary and passed through a multi-function compute unit.
Element-wise operations include addition and multiplication with a constant, allowing for the scaling or normalization of vectors.
\lb{Reduction} operations include both arithmetic and logical reductions.
An in-stream unit operating on 64-bit integers increases \gls{dma} engine area by only \SI{80}{\kGE} while accelerating element-wise scaling by up to \SI{32}{\x} and arithmetic reductions by up to \SI{12}{\x}.

\paragraph*{Packed irregular streams} We extend the cluster \gls{dma} engines and HBM3 endpoints with irregular stream \emph{packing} and \emph{unpacking} support.
This enables strided and indexed access streams to pack tightly into wide \gls{noc} flits, significantly improving link utilization and effective throughput on irregular workloads involving narrow HBM accesses.
The cluster-side extension provides scatter-gather \gls{dma} jobs, using index arrays in the cluster \gls{spm}. 
These jobs generate parallel streams of narrow ($\leq$64-bit) requests, up to eight of which are packed into one wide flit.
The HBM-side extension then unpacks these requests and combines them into minimum-size (256-bit) accesses using a temporal coalescer.
A \SI{161}{\kGE} packing unit improves the bandwidth efficiency of random-index scatter-gather by \SI{4.8}{\x}, ideally enabling up to \SI{8}{\x} improvements depending on the access pattern.

\subsection{Evaluation and SoA Comparison}
\label{sec:sogo:eval}

We estimate the area and timing of \sogo's compute chiplets in TSMC's \SI{7}{\nano\meter} FinFET technology based on a full cluster implementation with \emph{Fusion Compiler}, \sflo's existing \gls{d2d} \gls{phy} design, and the HBM3E interfaces of Nvidia's \emph{Blackwell} (B200) \gls{gpu}~\cite{nvidia2024blackwell}.
Despite their significant performance benefits presented in \cref{sec:sogo:arch}, we find that the combined area impact of our data movement extensions is only \SI{5.3}{\percent}.
Overall, we estimate that \sogo's chiplets are each \SI{112}{\milli\meter^2} in area and can operate at up to \SI{1.26}{\giga\hertz} under typical conditions; 
this yields a peak performance of \SI{10.3}{DP\dash\tera\flop\per\second} and a peak compute density of \SI{41.1}{DP\dash\giga\flop\per\second\per\milli\meter\squared}.

\newcommand{\tdl}[2]{\makecell[cc]{#1 \\ #2}}
\newcommand{\hdl}[2]{\makecell[lc]{#1 \\ #2}}
\newcommand{\ttl}[3]{\makecell[cc]{#1 \\ #2 \\ #3}}
\newcommand{\lsh}{\,/\,}

\begin{table}[t]
    \centering
    \caption{Evolution of our 2.5D manycore designs and comparison to Nvidia's \gls{soa} B200 \gls{gpu}, whose performance we approach.}
    \label{tab:soa_comparison}
    \vspace{-0.7mm}
    \renewcommand{\arraystretch}{1.1}
    \resizebox{\columnwidth}{!}{%
    \begin{threeparttable}
        \begin{tabular}{@{}lccc|c@{}}
            \hline
            \textbf{Metric} &
            \textbf{\socc~\cite{scheffler2025occamy}} &
            \textbf{\sflo~\cite{floonoc}} &
            \textbf{\sogo} &
            \textbf{B200~\cite{nvidia2024blackwell}} \\
            \hline
            Technology{\lsh}Die Count &
            GF\,12LP+ &
            GF\,12LP+ &
            TSMC\,N7&
            TSMC\,N4P\\
            \arrayrulecolor{ieee-dark-black-40}\hline
            Die{\lsh}Comp. Area [\si{\milli\meter\squared}] &
            146{\lsh}83.7 &
            143{\lsh}86.4 &
            448{\lsh}251 &
            1600{\lsh}737\tnote{a} \\
            \arrayrulecolor{ieee-dark-black-40}\hline
            Peak Perf. [\si{DP\dash\tera\flop\per\second}] &
            0.88 &
            1.29 &
            10.3 &
            40.0 \\
            \arrayrulecolor{ieee-dark-black-40}\hline
            \hdl{Peak Compute Density}{\text{[}\si{DP\dash\giga\flop\per\second\per\milli\metre\squared}]} &
            \tdl{10.5}{23.1\tnote{b}} &
            \tdl{14.9}{33.0\tnote{b}} &
            \tdl{41.1}{41.1} &
            \tdl{54.2\tnote{a}}{34.4\tnote{a,b}} \\
            \arrayrulecolor{ieee-dark-black-40}\hline
            En. Eff. \,\text{[}\si{DP\dash\giga\flop\per\second\per\watt}] &
            39.8\tnote{c} &
            42.2\tnote{c,d} &
            64.9\tnote{c,d} &
            82.1\tnote{e}~\,\cite{dongarra2025bwhpl} \\
            \arrayrulecolor{ieee-dark-black-40}\hline
            HBMx Configuration &
            2$\times$\,2E &
            2$\times$\,2E &
            4$\times$\,3 &
            8$\times$\,3E \\
            \arrayrulecolor{ieee-dark-black-40}\hline
            Total HBM BW [\si{\tera\byte\per\second}] &
            0.82 &
            0.82 &
            3.28 &
            8.00 \\
            \arrayrulecolor{ieee-dark-black-40}\hline
            Total D2D BW [\si{\tera\bit\per\second}] &
            0.07 &
            1.31 &
            8.44 &
            14.4 \\
            \hline
        \end{tabular}
        \vspace{-0.2mm}
        \begin{tablenotes}[para, flushleft]
        \item[a] Based on publicly available B200 render.
        \item[b] Normalized to TSMC N7 technology.
        \item[c] For FP64 \gls{gemm}.
        \item[d] Based on Occamy results and cluster, \gls{noc} power simulations.
        \item[d] For FP64 High-Perf. LINPACK; leverages FP emulation methods as described in~\cite{dongarra2025bwhpl}.
        \end{tablenotes}
    \end{threeparttable}
    }
\end{table}

\cref{tab:soa_comparison} summarizes the evolution of our open-source chiplet-based designs from {\socc} to {\sogo} and quantitatively compares them to Nvidia's \gls{soa} B200 GPU~\cite{nvidia2024blackwell}\cut{, which we aim to narrow the performance gap to}.
To provide additional insight, we estimate \sflo's and \sogo's FP64 \gls{gemm} energy efficiency based on {\socc}'s silicon measurements and cluster and \gls{noc} power simulations.
\sogo's peak performance is \SI{8.0}{\x} and \SI{11.8}{\x} higher than that of {\sflo} and {\socc}, achieving a \SI{2.8}{\x} and \SI{3.9}{\x} higher compute density and a \SI{54}{\percent} and \SI{63}{\percent}  higher energy efficiency, respectively.
While {\sogo} cannot match the peak performance of B200's \SI{3.6}{\x} larger dies, its compute density and energy efficiency are only \SI{24}{\percent} and \SI{21}{\percent} lower, despite its less advanced technology node (TSMC N7 vs. N4P).
In fact, when normalized to its \SI{7}{\nano\meter} node, {\sogo}'s compute density \emph{exceeds} that of B200 by \SI{19}{\percent}.
Its total HBM and \gls{d2d} bandwidths are \SI{4.0}{\x} and \SI{6.4}{\x} higher than those of {\sflo} and comparable to those of B200.

With \sogo~now exceeding the node-normalized compute density of an \gls{soa} \gls{gpu}, closing the remaining performance gap is largely a matter of increasing absolute die size.
We therefore shift our attention to increasing the degree of \emph{openness} of our 2.5D designs.

\section{Toward End-to-End Open-Source Chiplets}
\label{sec:e2ec}

So far, our chiplet-based designs have primarily offered synthesizable, open-source \gls{rtl} descriptions of their logic cores. 
However, making the source of subsequent design stages freely available would further enhance the benefits of our open-source hardware, promoting greater design transparency, reducing integration costs, and lowering barriers to collaboration between organizations. 
An \emph{end-to-end} open-source design, which can be reproduced without \emph{any} proprietary tools or \gls{ip}, would even enable zero-trust, step-by-step verification by third parties~\cite{sauter2025basilisk}. 
Therefore, in this section, we discuss different avenues to further expand the openness of our 2.5D designs.

\paragraph*{Simulation} Currently, our designs still rely on commercial simulators for their robustness and feature completeness.
Open-source simulators like \emph{Verilator}~%
and \emph{Icarus Verilog}~%
exist, but are limited by their scalability and lack of support for widespread SystemVerilog, testbench, and gate-level constructs.
However, they are rapidly improving: basic \gls{rtl} simulation of our logic core is already possible in Verilator, with support for advanced verification features and gate-level constructs like \gls{udp} being under active development.
These improvements may soon enable full \gls{rtl} and gate-level simulation of our designs using open-source simulators.

\paragraph*{EDA} Our implementations so far used commercial \gls{eda} tools, which clearly outperform their open-source counterparts.
However, logic synthesis with \emph{Yosys}~%
and place-and-route with \emph{OpenROAD}~%
are continuously improving their feature set and \gls{qor}, gradually closing this gap~\cite{sauter2025basilisk}.
We have already demonstrated our \emph{Cheshire} host platform in an end-to-end open-source \SI{130}{\nano\meter} \gls{soc}~\cite{sauter2025basilisk} and are currently implementing our cluster architecture in a \SI{22}{\nano\meter} node using open \gls{eda} tools~\cite{bertuletti2025opensource}.
Thus, implementing our full logic core in mature technology nodes using only open \gls{eda} may soon be possible.

\paragraph*{PDKs} Multiple open-source \glspl{pdk} for manufacturable technologies, along with open cell libraries, have been released in recent years.
This includes GlobalFoundries' \SI{180}{\nano\meter}, SkyWater's \SI{130}{\nano\meter}, and IHP's \SI{130}{\nano\meter} \glspl{pdk}, \lb{as well as ICsprout's recently released \SI{55}{\nano\meter} \gls{pdk}}.
We have already demonstrated our host subsystem in IHP's open technology~\cite{sauter2025basilisk}.
However, these mature $\geq$\SI{55}{\nano\meter} nodes are insufficient to produce high-complexity designs like our $\leq$\SI{12}{\nano\meter} compute chiplets.
This cannot be remedied by designers themselves, but only by advanced foundries choosing to open-source their \glspl{pdk}.

\paragraph*{Off-die \glspl{phy}} High-speed \glspl{phy} for DRAM and \gls{d2d} interfaces may present the biggest hurdle to expanding openness,
as their mixed-signal nature requires expert design and is strongly coupled to the \gls{eda} tools used and \gls{pdk} targeted.
Existing open-source \glspl{soc} often work around this by falling back to slower, fully-digital interfaces~\cite{ottaviano2023cheshire,sauter2025basilisk}.
However, mixed-signal design for open \glspl{pdk} \emph{is} possible in principle, as demonstrated by Efabless' analog design flow~\cite{edwards2023efabless}.
Thus, open-source \glspl{phy} \lb{could} eventually become available, albeit subject to the performance limitations of open-source \gls{eda} tools and \glspl{pdk}.

In summary, as open-source simulation and EDA tools gradually approach SoA quality, the lack of advanced open PDKs and the inherent difficulty of open PHY design are emerging as the primary challenges to further increasing the openness of our chiplet designs.

\section{Conclusion}

In this paper,
we presented our evolving roadmap for open-source, chiplet-based RISC-V manycores targeting \gls{hpc} and \gls{ai} workloads and aiming to close the performance gap to commercial silicon.
We began with \emph{\socc}, the first silicon-proven dual-chiplet, dual-HBM2E RISC-V manycore in \SI{12}{\nano\meter} FinFET,  establishing an \SI{876}{DP\dash\giga\flop\per\second} baseline with competitive (\SI{89}{\percent}) and leading (42--\SI{83}{\percent}) peak \gls{fpu} utilizations on regular and irregular workloads, respectively, and thereby proving the efficacy of our cluster-based compute architecture.
With \emph{\sflo}, we then upgraded \socc's crossbar-based interconnect to a scalable \SI{0.15}{\pico\joule\per\byte\per hop} 2D mesh NoC, raising peak performance to \SI{1.29}{DP\dash\tera\flop\per\second} without increasing chiplet area and improving compute density, high-load HBM bandwidth, and  wide-transfer \gls{d2d} bandwidth by \SI{43}{\percent}, \SI{22}{\percent}, and \SI{16}{\x}, respectively.
Finally, we proposed \emph{\sogo}, a \SI{7}{\nano\meter} quad-chiplet, quad-HBM3 concept architecture scaling \sflo's compute mesh to \SI{10.3}{DP\dash\tera\flop\per\second} and introducing lightweight extensions for in-router collectives, in-stream \gls{dma} operations, and packed irregular streams.
{\sogo} exceeds the node-normalized compute density of Nvidia's \gls{soa} B200 \gls{gpu} by \SI{19}{\percent} while achieving a comparable energy efficiency and off-die bandwidths, reducing the remaining performance gap to a matter of die size.
\lb{Our analysis demonstrates that open-source  2.5D designs can scale to \gls{soa} performance in their digital core.}
\lb{If we raise our ambition toward end-to-end open-source chiplets, we identify the lack of open \glspl{pdk} for advanced technologies and the challenges in open \gls{phy} design as the main bottlenecks to be addressed in the future}.

%

%

\renewcommand{\baselinestretch}{\refbls}


\end{document}